%% file: 0_main.tex
\documentclass[conference]{IEEEtran}
\IEEEoverridecommandlockouts

\usepackage{algorithmic}
\usepackage[linesnumbered,ruled,vlined]{algorithm2e}
\usepackage{multirow}
\usepackage{multicol}
\usepackage{wrapfig}
\usepackage{soul}
\usepackage{balance}

\usepackage{caption}
\usepackage{subcaption}
\usepackage{graphicx} 
\usepackage{subcaption} 
\usepackage{subfig}

\usepackage{epstopdf}

\usepackage{amsfonts}
\usepackage[flushleft]{threeparttable}
\usepackage{adjustbox}
\usepackage{fancyhdr}
\usepackage[dvipsnames]{xcolor}
\usepackage[flushleft]{threeparttable}
\usepackage{tikz}
\usepackage{outlines}
\usepackage{tcolorbox}
\usepackage{blindtext}
\usepackage{xcolor}
\usepackage{color}
\usepackage{listings}
\usepackage{color}
\usepackage{MnSymbol,wasysym}
\usepackage[switch]{lineno}
\usepackage{array}

\usepackage{amsmath,amsthm}

\usepackage{mathtools} 

\usepackage[mathscr]{euscript}
\usepackage{caption} 

\usepackage{xcolor,colortbl}
\def\BibTeX{{\rm B\kern-.05em{\sc i\kern-.025em b}\kern-.08em
    T\kern-.1667em\lower.7ex\hbox{E}\kern-.125emX}}

\usepackage{tikz}

\usepackage[T1]{fontenc} 

\usepackage{subcaption}

\theoremstyle{plain}

\begin{document}

\title{System-Level Performance Modeling of \\
Photonic In-Memory Computing}


\author{
    \IEEEauthorblockN{Jebacyril Arockiaraj\IEEEauthorrefmark{2}, Sasindu Wijeratne\IEEEauthorrefmark{2}, Sugeet Sunder\IEEEauthorrefmark{1}, Md Abdullah-Al Kaiser\IEEEauthorrefmark{3} \\ Akhilesh Jaiswal\IEEEauthorrefmark{3}, Ajey P. Jacob\IEEEauthorrefmark{1}, Viktor Prasanna\IEEEauthorrefmark{2}}
    \IEEEauthorblockA{\IEEEauthorrefmark{2}Ming Hsieh Department of Electrical and Computer Engineering, University of Southern California}
    \IEEEauthorblockA{\IEEEauthorrefmark{1}Information Sciences Institute (ISI), University of Southern California}
    \IEEEauthorblockA{\IEEEauthorrefmark{3}Electrical and Computer Engineering, University of Wisconsin-Madison}
    Email: \{arockiar, kangaram, prasanna\}@usc.edu, \{sunder, ajey\}@isi.edu, \{mkaiser8, akhilesh.jaiswal\}@wisc.edu
}

\maketitle
\begin{abstract}
Photonic in-memory computing is a high-speed, low-energy alternative to traditional transistor-based digital computing that utilizes high photonic operating frequencies and bandwidths. In this work, we develop a comprehensive system-level performance model for photonic in-memory computing, capturing the effects of key latency sources such as external memory access and opto-electronic conversion. We perform algorithm-to-hardware mapping across a range of workloads, including the Sod shock tube problem, Matricized Tensor Times Khatri-Rao Product (MTTKRP), and the Vlasov-Maxwell equation, to evaluate how the latencies impact real-world high-performance computing workloads. Our performance model shows that, while accounting for system overheads, a compact 1$\times$256 bit single-wavelength photonic SRAM array, fabricated using the standard silicon photonics process by GlobalFoundries, sustains up to 1.5 TOPS, 0.9 TOPS, and 1.3 TOPS on the Sod shock tube problem, MTTKRP, and the Vlasov-Maxwell equation with an average energy efficiency of 2.5 TOPS/W.
\end{abstract}

\begin{IEEEkeywords}
Photonic Computing, System-Level Modeling, Sod Shock Tube Problem, MTTKRP, Vlasov–Maxwell Equation
\vspace{-3mm}
\end{IEEEkeywords}

\input{1_Introduction}

\input{2_Background}
\input{3_Applications}
\input{4_System_Performance_Model}
\input{5_Streaming_Algorithm}
\input{6_Results}

\input{7_Conclusion}

\section*{ACKNOWLEDGEMENT}
This work is supported by the Defense Advanced Research Projects Agency (DARPA) under grant N660012424003. We would like to thank Mukund Vengalattore and his team at DARPA for their support. We also thank Justin Ray Angus and his team at Lawrence Livermore National Laboratory (LLNL) for their input and valuable discussions.
\bibliographystyle{IEEEtran}
\bibliography{reference}

\end{document}

%% file: 1_Introduction.tex
\vspace{-1mm}
\section{Introduction}
\vspace{-1mm}
As CMOS technology nears its scaling limits, traditional methods, such as parallel processing, reach bottlenecks in latency and energy efficiency due to scalar computation, high resistance, and capacitive coupling in electrical interconnects~\cite{Cho2022SRAM,9365766,9250531}. Emerging technologies, such as optical computing~\cite{acsphotonics}, offer a path forward by overcoming these limitations of the traditional systems~\cite {psram_dac}.

In optical computing, photonic static random-access memory (pSRAM)~\cite{psram} offers a promising alternative to traditional memory technologies, delivering ultra-low energy consumption and significantly higher bandwidth through the use of compatible optical components~\cite{osram_arxiv}. Beyond data storage, pSRAM arrays also serve as compute cores by leveraging optical pulses for in-memory computation~\cite{sunder2025scalable}. While prior works~\cite{10938433, 9926291, 9926407} have explored device-level performance, a key gap remains: the lack of comprehensive system-level performance models that account for factors such as external memory access and opto-electronic conversion, which is essential to evaluate whether device-level advantages persist at the system-level and to guide architectural design.

In this work, we model a pSRAM array architecture as a three-part system: pSRAM for in-memory compute, electrical external memory, and an opto-electronic interface. These components are identified as the primary contributors to system latency, forming the basis of our system-level performance model. To support systematic algorithm-to-hardware mapping, we introduce a network model abstraction of the pSRAM array, which encapsulates hardware capabilities through well-defined computational primitives. 
We develop streaming algorithms for three diverse high-performance computing workloads: the Sod shock tube problem, MTTKRP, and the Vlasov–Maxwell equation. We map the core computational kernels of these workloads to the network model. The streaming algorithms operate without intermediate optical storage to model a conservative performance baseline. Although we demonstrate this approach using three representative workloads, any algorithm that can be expressed using these primitives can be mapped onto the pSRAM architecture.

The Sod shock tube problem~\cite{SOD19781} is a standard benchmark for evaluating numerical solvers of the Euler equation, which are fundamental in plasma physics~\cite{llnl}. Its applications span manufacturing and materials processing~\cite{chen}, inertial confinement fusion~\cite{llnl}, and biomedical diagnostics and therapeutics~\cite{laroussi}. Matricized Tensor Times Khatri-Rao Product (MTTKRP) is a computationally expensive kernel in the tensor decomposition~\cite{07070111X}, used in machine learning~\cite{rabanser2017}, signal processing~\cite{7891546}, and data mining~\cite{4781131}. The Vlasov–Maxwell equation models the evolution of charged-particle distributions under electromagnetic fields and remains a fundamental challenge in kinetic plasma simulations~\cite{CROUSEILLES2015224}.

In this paper, we introduce a comprehensive system-level performance model for photonic in-memory-based computing architectures as part of the ongoing Scalable Photonic SRAM-based in-memory computing tensor core (SPRINT) project~\cite{sunder2025scalable}.
The main contributions of this work are as follows:
\begin{itemize}
    \item We develop a system-level performance model for pSRAM, capturing latency attributions from pSRAM, external memory, and opto-electronic conversion.
    \item We define a network model abstraction of the pSRAM array and develop streaming algorithms for three representative workloads: the 1D Sod shock tube problem, MTTKRP, and the Vlasov-Maxwell equation, allowing systematic mapping of algorithms to the pSRAM.
    \item We use the performance model to evaluate architectural trade-offs across bandwidth, frequency, array size, and conversion latency, and identify compute- and memory-bound regimes through roofline analysis.
    \item Our analysis indicates that a compact 1$\times$256 bit pSRAM array with system overheads such as external memory access and opto-electronic conversion, sustains up to 1.5 TOPS, 0.9 TOPS, and 1.3 TOPS on the Sod shock tube problem, MTTKRP, and the Vlasov-Maxwell equation, with a pSRAM energy efficiency of 2.5 TOPS/W.
\end{itemize}

%% file: 2_Background.tex
\section{Photonic SRAM Architecture} \label{optical_sram}

In this section, we briefly introduce the pSRAM architecture as the foundation for the performance modeling and algorithm mapping described in later sections. The pSRAM array is organized as a grid of photonic bitcells with integrated multiply-accumulate functionality~\cite{sunder2025scalable}. For a target bit precision of $w$, each compute cell consists of $w$ bitcells. We define $P$ as the total number of compute cells in the pSRAM array. The following section describes how a single compute cell composed of $w$ bitcells performs in-memory computation.

\begin{figure}[!ht]
\vspace{-1mm}
\centering
\includegraphics[width=1\linewidth]{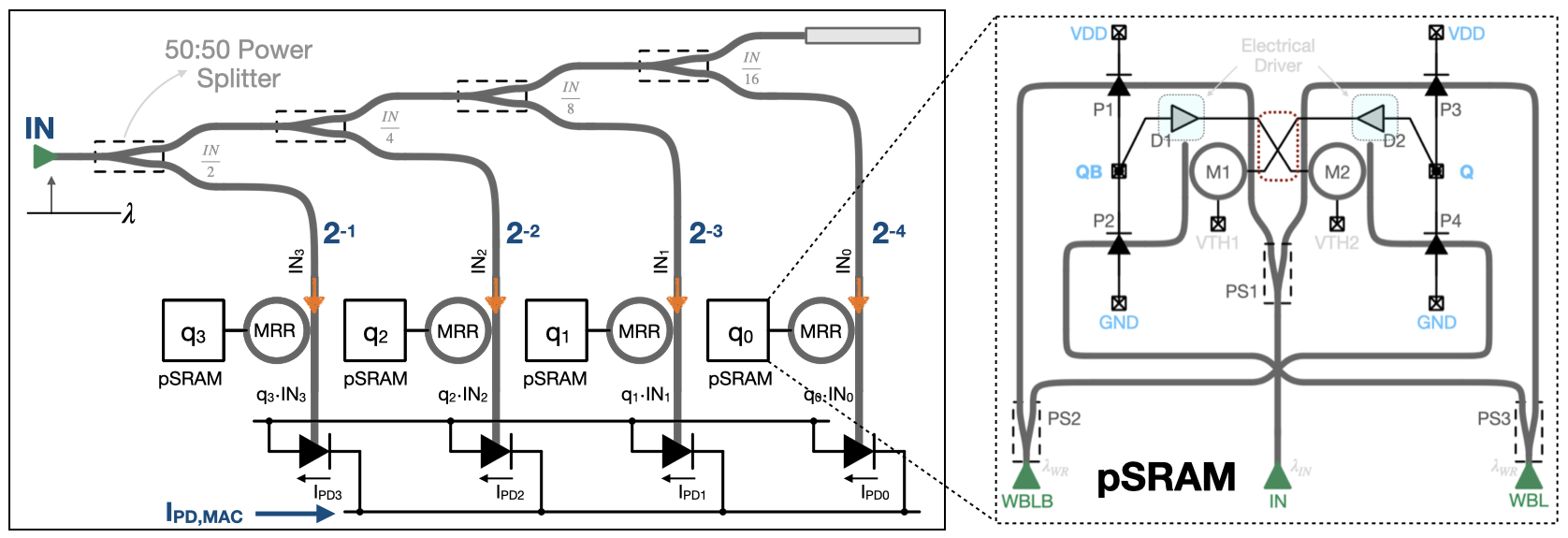}
\caption{Schematic of a 4-bit $(w = 4)$ mixed-signal pSRAM compute cell with inputs scaled based on their bit significance and weight bits stored in the pSRAM bitcells.}
\label{fig_hardware}
\vspace{-1mm}
\end{figure}

\subsubsection{Input Encoding}

The input signal to the pSRAM bitcell is modulated using an electro-optic intensity modulator.
This intensity-modulated signal carries the encoded data sequentially through time, avoiding the need for complex wavelength shaping components. 
\vspace{1mm}
\subsubsection{Bitcell}

The pSRAM bitcell consists of a cross-coupled structure based on microring resonators (M1 and M2) and balanced photodiodes (P1, P2, P3, and P4), which store differential binary data on optically controlled electrical nodes W and WB, as depicted in Figure~\ref{fig_hardware}. The pSRAM bitcell maintains data in a static state under continuous optical and electrical bias~\cite{psram}. It also supports high-speed optical write operations through differential write bitlines (WBL and WBLB). With a -20 dBm optical bias and a wall-plug efficiency of 0.23 \cite{wpe}, the pSRAM bitcell consumes 0.5 pJ of energy per switching event while operating at a speed of 20 GHz \cite{psram_dac}.

The architecture has been fabricated using the GlobalFoundries' 45SPCLO PDK \cite{rakowski202045nm}, ensuring compatibility with CMOS-based electronic subsystems, and its hold and write functionality has been experimentally validated~\cite{psram_dac}.
\vspace{1mm}
\subsubsection{Output Encoding}

Each pSRAM bitcell stores binary data. Scaled input light intensities, adjusted for bit significance, propagate along the word lines and interact with $w$ pSRAM bitcells per compute cell, each storing a corresponding weight bit as shown in Figure~\ref{fig_hardware}. This enables bitwise multiplication of inputs and stored weights, with the resulting partial dot products computed by ring modulators. These partial products are then aggregated by a peripheral photodiode array to produce the final multi-bit multiply-accumulate (MAC) output \cite{psram_dac, sunder2025scalable}.
This single-wavelength approach remains compatible with integrated photonics and offers a scalable path for optical memory and compute systems in hybrid photonic-electronic architectures.

%% file: 3_Applications.tex
\section{Target Applications}

State-of-the-art computing research increasingly focuses on high-performance scientific applications, especially in areas such as turbulent flows, plasma dynamics, and heat transport in heterogeneous materials~\cite{darpa_napsac}. Reflecting these priorities, we evaluate two scientific applications in this work: the 1D Sod shock tube problem and the Vlasov–Maxwell equation. Additionally, to address the growing importance of data-driven workloads, we evaluate MTTKRP, a key kernel in machine learning and signal processing.

\subsection{Sod Shock Tube: Plasma Physics Benchmark}~\label{1dsst_intro}\vspace{-5mm}


In the 1D Sod shock tube problem numerical solution (1D SST-NS), the Euler equation is discretized on a 1D grid. The algorithm updates the solution vector through three steps: (i) flux computation across cell interfaces, (ii) predictor step using flux differences to update cell centers, and (iii) corrector step that finalizes the update at each grid point~\cite{llnl}. The update equations are as follows:
    \begin{equation}
    F_i^t = F_{i-1/2}^t + F_{i+1/2}^t + j_{i-1/2}^t W_{i-1/2}^t - j_{i+1/2}^t W_{i+1/2}^t
    \end{equation}

    \begin{equation}
    W_{i+1/2}^{t+1/2} = W_{i+1/2}^t - k \left(F_{i+1}^t - F_i^t\right)
    \end{equation}

    \begin{equation}
    W_{i+1/2}^{t+1} = W_{i+1/2}^t - 2k \left(F_{i+1}^{t+1/2} - F_i^{t+1/2}\right)
    \end{equation}

where $W$ is the solution vector, $F$ represents the flux vector, $j$ is the maximum eigenvalue of the flux Jacobian, $i$ is the index of the grid point, $k=\frac{\Delta t}{4 \Delta x}$ is a constant based on $\Delta t$ the time step, and $\Delta x$ the grid spacing.



\subsection{MTTKRP: Canonical Polyadic Tensor Decomposition}~\label{mttkrp_intro}
Canonical Polyadic Decomposition (CPD) decomposes $\mathcal{X}$ into a sum of rank-1 tensors, which best approximates $\mathcal{X}$~\cite{07070111X}. For example, given the 3-mode tensor $\mathcal{X} \in \mathbb{R}^{I_0 \times I_1 \times I_2}$, our goal is to approximate the original tensor as \vspace{-2mm}
\begin{equation} \label{eqn_approx_tensor}
\begin{split}
\mathcal{X} \approx \sum_{i=0}^{R-1} \mathbf{a}_i \circ \mathbf{b}_i \circ \mathbf{c}_i
\end{split}
\end{equation}
\vspace{-3mm}

where $R$ is a positive integer and $\mathbf{a}_i \in \mathbb{R}^{I_0}$, $\mathbf{b}_i \in \mathbb{R}^{I_1}$, and $\mathbf{c}_i \in \mathbb{R}^{I_2}$. The alternating least squares (ALS) method is used to compute CPD, where MTTKRP is iteratively performed on all the Matricizations of $\mathcal{X}$~\cite{07070111X}. In this paper, performing MTTKRP on all the Matricizations of an input tensor is called computing MTTKRP along all the modes. The outputs $\mathbf{A}$, $\mathbf{B}$, and $\mathbf{C}$ are the factor matrices that approximate $\mathcal{X}$. $\mathbf{a}_i$, $\mathbf{b}_i$, and $\mathbf{c}_i$ in Equation~\ref{eqn_approx_tensor} refers to the $i^{\text{th}}$ column of $\mathbf{A}$, $\mathbf{B}$, and $\mathbf{C}$, respectively.



\subsection{Vlasov–Maxwell Equation}~\label{Vlasov_intro}
The core computational kernel in the spectral Vlasov–Maxwell approach is the convolution in Fourier space, which is efficiently implemented using Fast Fourier Transforms (FFT)~\cite{DELZANNO2015338,10.1063}. 
\begin{equation}
H * C = \mathrm{IFFT} \left[ \mathrm{FFT}(H) \times \mathrm{FFT}(C) \right]
\label{eq:fft_conv}
\end{equation}

where $H$ and $C$ denote intermediate coefficient arrays of the distribution function. This process is primarily composed of elementwise complex multiplication~\cite{10.5555/47314,10.5555/294797}, making it the dominant arithmetic operation in spectral Vlasov–Maxwell solvers.

%% file: 4_System_Performance_Model.tex
\section{System-Level Performance Model}~\label{performance_model}

In this section, we develop a comprehensive system-level performance model for the pSRAM-based in-memory computing architecture by identifying the primary latency components that impact performance and the resulting sustained throughput of the system.

\subsection{Hardware Modules}

\begin{figure}[ht]
\vspace{-3mm}
\centering
\includegraphics[width=0.9\linewidth]{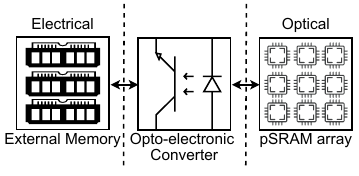}
\vspace{-2mm}
\caption{Overall system illustrating the interaction between external memory (electrical domain), the opto-electronic converter, and the pSRAM array (optical domain).}
\label{fig:system}
\vspace{-3mm}
\end{figure}

The overall system, illustrated in Figure~\ref{fig:system}, consists of three primary components: a pSRAM array, which comprises multiple compute cells and serves as the photonic compute core; an external memory unit operating in the electrical domain for data storage; and an opto-electronic converter that facilitates the data exchange between the two domains. 

\subsection{System Performance Modeling}
In this section, we present a detailed analytical formulation of the system-level performance model. This model quantifies the end-to-end latency incurred during the execution of a compute workload on the pSRAM-based system. The compute workload consists of $N_\text{total}$ computations, where each computation corresponds to a basic arithmetic operation (e.g., addition). The key parameters of the model are:

 \begin{itemize}
    \item $N_\text{total}$: Total number of computations in the compute workload.
    \item $S$: Total size of input and output data transferred per compute workload (in bits), corresponding to external memory.
    \item $B$: Peak external memory bandwidth (in bits/second).
    \item $T_\text{access}$: Total external memory access latency.
    \item $T_{\text{conv}}$: Total opto-electronic conversion latency, including electrical-to-optical and optical-to-electrical latencies.
    \item $T_{\text{comp}}$: Time spent performing computation using the pSRAM array.
    \item $P$: Number of compute cells in the pSRAM array.
    \item $F$: pSRAM operating frequency (in Hz).
    \item $Ops$: Number of operations performed per cycle by each compute cell.
\end{itemize}

The total time to execute the compute workload is:
\begin{equation}
T_{\text{total}} = T_{\text{mem}} + T_{\text{conv}} + T_{\text{comp}}
\label{eqn6_o}
\end{equation}

Each term in Equation~\ref{eqn6_o} is defined as follows.

\textbf{External Memory Access:}
Data transfer between external memory and the converter incurs both a fixed memory access latency (e.g., DRAM row access) and a transfer time that depends on the electrical interconnect bandwidth, which we model as:
\begin{equation}
T_{\text{mem}} = T_{\text{access}} + \frac{S}{B}
\end{equation}

\textbf{Opto-Electronic Conversion:}
The opto-electronic converter introduces a fixed latency due to signal conversion in both directions. While these conversions can be pipelined, the initial conversion latency contributes to the end-to-end latency. 
\begin{equation}\label{opto}
T_{\text{conv}} = T_{\text{EO}} + T_{\text{OE}}
\end{equation}
where $T_{\text{EO}}$ and $T_{\text{OE}}$ denote fixed latencies for electrical-to-optical and optical-to-electrical conversion, respectively. These latencies can be partially overlapped in pipelined execution.

\textbf{pSRAM Computation:}
Once the input signals are available in the optical domain, the pSRAM compute cells perform the operations with a fixed computation latency per operation in parallel. 
\begin{equation}
T_{\text{comp}} = \frac{N_\text{total}}{P \times Ops\times F}
\end{equation}

\subsection{Sustained Performance}
The sustained performance (in operations per second) is derived from $T_{\text{total}}$, the total time required to execute the compute workload:
\vspace{-3mm}
\begin{equation}
\text{Sustained Performance} = \frac{N_\text{total}}{T_{\text{total}}}
\end{equation}
\begin{equation}
    T_{\text{total}} = T_{\text{access}} + \frac{S}{B} + T_{\text{conv}} + \frac{N_\text{total}}{P \times Ops\times F}
\end{equation}

This formulation captures how architectural parameters, such as memory bandwidth ($B$) and conversion latency ($T_{\text{conv}}$), impact sustained throughput. For example, increasing memory bandwidth or reducing conversion latency directly improves sustained performance.

\subsection{Peak Performance}

For completeness, we define the peak performance, representing the theoretical upper bound under ideal conditions with no data transfer or conversion overhead. The peak performance, which captures the raw compute capability of the pSRAM array, is given by:

\begin{equation}
    \text{Peak performance} = P \times F \times Ops
\end{equation}

\subsection{Impact of Bit Width on Performance}
The number of compute cells $P$ that can be formed within the pSRAM array is determined by the total available storage capacity in the pSRAM array and the input bit width, which reflects the precision of the data. 

This relationship can be expressed as:
\begin{equation}
P = \frac{C_{\text{total}}}{w}
\end{equation}

where $C_{\text{total}}$ is the total number of bits available in the pSRAM array, and $w$ is the bit width of each operand. Reducing the bit width allows more compute cells to be formed, thereby increasing parallelism and performance. However, this comes at the cost of reduced precision. Therefore, the choice of input bit width must balance the trade-off between performance and application-specific precision requirements.

%% file: 5_Streaming_Algorithm.tex
\section{Mapping Applications to\\ System-Level Performance Model} 
To evaluate how different applications execute on the pSRAM, we introduce a network model of the pSRAM array that enables structured algorithm-to-hardware mapping. Leveraging this model, we develop streaming algorithms, in which input data is fetched from external memory and results are written back without any intermediate optical storage. These algorithms impose stringent demands on external memory bandwidth and data movement, thereby effectively modeling a conservative worst-case performance baseline. In practice, architectural enhancements such as optical buffering can reduce bandwidth bottlenecks by enabling data reuse and better scheduling of data transfers, leading to substantial performance improvements beyond this lower bound. This makes streaming algorithms a conservative yet valuable tool for architectural benchmarking.

\subsection{Network Model}\label{network_model}
We define a high-level, synchronous network abstraction model for algorithm-hardware co-design, where each compute cell performs a local operation and exchanges data with its immediate neighbors within a single time step. In this work, we consider an $M$-processor network model consisting of $M$ photonic compute cells interconnected in a 1D mesh topology with $M=N$, where $N$ denotes the total number of iteration points per loop iteration in an algorithm. For example, $N$ corresponds to the number of grid points in 1D SST-NS described in Section~\ref{1dsst_intro}.

We define two key primitives in this model: computation and communication.

\noindent\textbf{Computation Primitive:}
Each compute cell supports a multiply-accumulate primitive:

\begin{itemize}
  \item \texttt{LocalMAC(op, a, b, c) $\rightarrow$ z}
  
  Performs either $z = c + ab$ or $z = c - ab$, depending on the operation type $\texttt{op} \in \{\texttt{add}, \texttt{sub}\}$.

  \item $a$ is a constant operand preloaded and stored in the pSRAM compute cell.

  \item $b$ and $c$ are input operands.
\end{itemize}

\noindent\textbf{Communication Primitives:}
Each compute cell can exchange scalar values with its immediate neighbors:

\begin{itemize}
  \item \texttt{SendToNeighbor(dir, data)}: Sends scalar \texttt{data} to the neighbor in direction \texttt{dir} $\in \{\texttt{left}, \texttt{right}\}$.
  
  \item \texttt{RecvFromNeighbor(dir) $\rightarrow$ data}: Receives the scalar \texttt{data} from the neighbor in direction \texttt{dir}.
\end{itemize}

\subsection{Streaming Algorithm for 1D SST-NS}

The 1D SST-NS consists of $N$ grid points, each of which is updated at every time step, with each grid point $i$ mapped to a dedicated compute cell in the 1D mesh network model. The streaming algorithm representing the operations performed by a single compute cell per time step is presented in Algorithm~\ref{alg:sst_streaming}. The notation of 1D SST-NS is discussed in Section~\ref{1dsst_intro}.

\vspace{-1mm}
\begin{algorithm}
\DontPrintSemicolon
\KwIn{Solution $w_i$, flux $f_i$, constants $j$, $k$}
\KwOut{Updated solution vector $w_i$}

$f_{R,i} \gets \texttt{LocalMAC}(\texttt{sub}, j, w_i, f_i)$ \;
$f_{L,i} \gets \texttt{LocalMAC}(\texttt{add}, j, w_i, f_i)$ \;

\texttt{SendToNeighbor}(left, $f_{R,i}$) \;
$f_{R,i+1} \gets \texttt{RecvFromNeighbor}(right)$ \;

$f_{\text{cell},i} \gets \texttt{LocalMAC}(\texttt{sub}, 1, f_{\text{cell},i}, f_{\text{cell},i-1})$

\texttt{SendToNeighbor}(right, $f_{\text{cell},i}$) \;
$f_{\text{cell},i-1} \gets \texttt{RecvFromNeighbor}(left)$ \;

$f_{\text{cell},i} \gets \texttt{LocalMAC}(\texttt{sub}, 1, f_{\text{cell},i}, f_{\text{cell},i-1})$ \;
$w_i \gets \texttt{LocalMAC}(\texttt{sub}, k, f_{\text{cell},i}, w_i)$ \;

\Return $w_i$ \;
\caption{Streaming Algorithm for 1D SST-NS}
\label{alg:sst_streaming}
\end{algorithm}
\vspace{-2mm}

\subsection{Streaming Algorithm for MTTKRP}

\vspace{-3mm}
\begin{algorithm}
\DontPrintSemicolon
\KwIn{A sparse tensor $\mathcal{X} \in \mathbb{R}^{I_0 \times I_1 \times I_2}$, dense factor matrices $\mathbf{{B}} \in \mathbb{R}^{I_1 \times R}$, $\mathbf{{C}} \in \mathbb{R}^{I_2 \times R}$}
\KwOut{Updated dense factor matrix $\mathbf{A} \in \mathbb{R}^{I_0 \times R}$}

\textcolor{blue}{// Step 1: Hadamard Product of Factor Matrix Rows} \;
\For{each $h_1$ factor matrix row in $\mathbf{B}$}{
\For{each $h_2$ factor matrix row in $\mathbf{C}$}{
$f(h_1 \times I_1 + h_2,i) \gets \texttt{LocalMAC}(\texttt{add}, B(h_1,i), C(h_2, i), 0)$ \;
}
}

\textcolor{blue}{// Step 2: Scaling with a Tensor Element} \;

\For{each $h_0$ output factor matrix row in $\mathbf{A}$}{
\For{each nonzero element in $\mathcal{X}$ at $(h_0,h_1,h_2)$ with $h_0$ coordinates}{
$\mathbf{A}(h_0, i) \gets \texttt{LocalMAC}(\texttt{add}, \mathcal{X}(h_0, h_1, h_2),$\;
$f(h_1 \times I_1 + h_2,i), \mathbf{A}(h_0, i))$ \;
}
}
\Return $A$ \;
\caption{Streaming Algorithm for Mode 0 MTTKRP of a 3D Tensor Using Network Model Primitives}
\label{alg:mttkrp_streaming}
\end{algorithm}
Algorithm~\ref{alg:mttkrp_streaming} shows the computation of mode 0 MTTKRP of a 3-mode tensor using the network model in streaming setting. For further details on MTTKRP and the computation pattern, refer to our previous work~\cite{10938433}. The algorithm consists of factor matrices with rank $R$, each assigned to a dedicated compute cell in the 1D mesh network model. The Algorithm~\ref{alg:mttkrp_streaming} shows the operations executed in the $i^{\text{th}}$ compute cell on the network. The notation of MTTKRP is discussed in Section~\ref{mttkrp_intro}.

\subsection{Streaming Algorithm for Vlasov–Maxwell Equation}
The streaming algorithm for the Vlasov-Maxwell equation is shown in Algorithm~\ref{alg:vm_streaming}. We formulate the streaming algorithm using network model primitives to capture the core computations in the spectral Vlasov–Maxwell solver. Each Fourier mode index is mapped to a dedicated compute cell in the 1D mesh. The compute cells receive two complex-valued inputs in the Fourier domain and perform the elementwise complex multiplication. The notation of the Vlasov-Maxwell equation is discussed in Section~\ref{Vlasov_intro}.
\vspace{-2mm}
\begin{algorithm}
\DontPrintSemicolon
\KwIn{Fourier mode $\hat{f}=f_R+jf_I$, constant $\hat{k}=k_R+jk_I$, accumulator $\hat{z}=z_R+j z_I$}
\KwOut{Updated Fourier mode $\hat{f}$}

$temp \gets \texttt{LocalMAC}(\texttt{add}, k_R, z_R, 0)$ \;
$temp \gets \texttt{LocalMAC}(\texttt{sub}, k_I, z_I, temp)$ \;
$f_R \gets \texttt{LocalMAC}(\texttt{add}, 1, temp, f_R)$ \;

$temp \gets \texttt{LocalMAC}(\texttt{add}, k_I, z_R, 0)$ \;
$temp \gets \texttt{LocalMAC}(\texttt{add}, k_R, z_I, temp)$ \;
$f_I \gets \texttt{LocalMAC}(\texttt{add}, 1, temp, f_I)$ \;

\Return $(f_R,\,f_I)$ \;
\caption{Streaming Algorithm for Spectral Vlasov--Maxwell Equation}
\label{alg:vm_streaming}
\end{algorithm}
\vspace{-3mm}

\subsection{Roofline Analysis of Streaming Workloads}
We construct a roofline model~\cite{roofline} of the pSRAM array using HBM3E~\cite{10536972} as external memory to provide a high-level, throughput-oriented performance view. This analysis reveals that the two scientific workloads are compute-bound, while MTTKRP is memory-bound. Notably, this classification is not fixed; core architectural parameters from our performance model significantly influence the system bottleneck.

\begin{figure}[!ht]
\vspace{-4mm}
\centering
\includegraphics[width=0.9\linewidth]{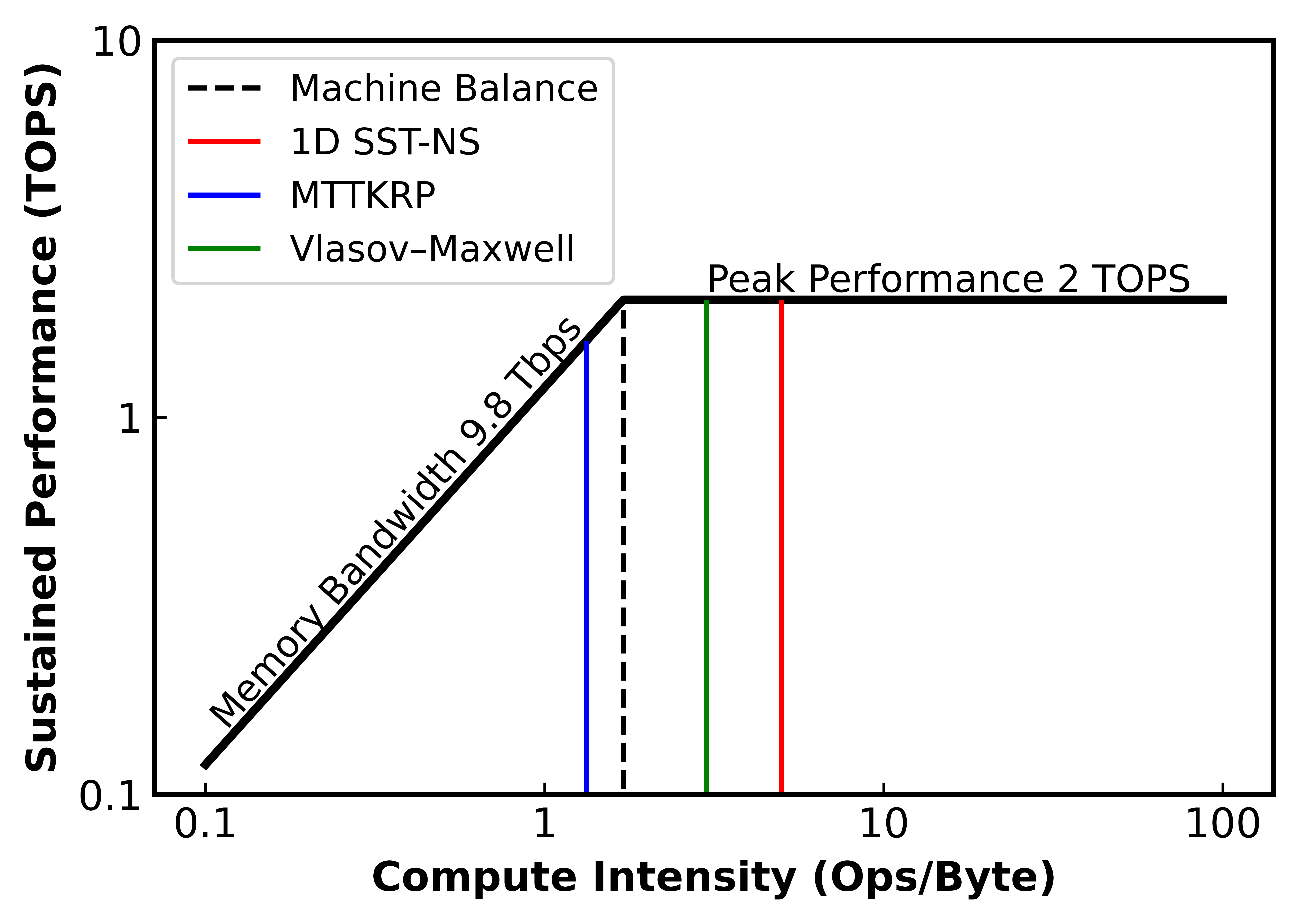}
\caption{Roofline performance model of the pSRAM array.}
\label{fig:roofline}
\vspace{-3mm}
\end{figure}

Reducing input bit width increases the number of operations performed per byte transferred from external memory, effectively shifting memory-bound workloads such as MTTKRP closer to the compute roof. Increasing peak external memory bandwidth shifts the machine balance to the left, enhancing performance in memory-bound workloads. Conversely, raising the operating frequency of the pSRAM array elevates the compute roof vertically, improving throughput for compute-intensive workloads. By visualizing the impact of core architectural parameters, the roofline model provides prescriptive guidance. 

\subsection{Mapping to pSRAM}

In Section~\ref{network_model}, we described an $M$-processor network model in which each compute cell is assigned a single iteration point. However, the pSRAM array contains $P$ compute cells, where $P<N$, total iteration points. Therefore, we apply a \textit{block distribution} strategy~\cite{grama03}, where compute cell $P_i$ is assigned a contiguous block in the range $[i \cdot N/P, (i+1) \cdot N/P - 1]$. This mapping ensures balanced workload distribution, with communication limited to the boundaries of each block. 


%% file: 6_Results.tex
\section{Results}~\label{experiments} \vspace{-8mm}

\subsection{Experimental Setup}
We select our performance model parameters based on prior work~\cite{psram_dac, sunder2025scalable}. Specifically, we consider a 1D pSRAM array with 1x256 bits operating at a frequency of $F=32$ GHz with a bit precision of 8, resulting in $P=32$ compute cells. We consider HBM3E~\cite{10536972} as the external memory. Each compute cell performs two operations and communicates with its neighboring cell in one clock cycle, resulting in $Ops=2$. 

\subsection{System Performance Analysis}
Using performance modeling, we evaluate the impact of three major latency components identified in Section~\ref{performance_model} on the sustained performance of the pSRAM array.
\vspace{3mm}
\subsubsection{Peak External Memory Bandwidth}
\begin{figure}[!ht]
\vspace{-2mm}
\centering
\includegraphics[width=0.9\linewidth]{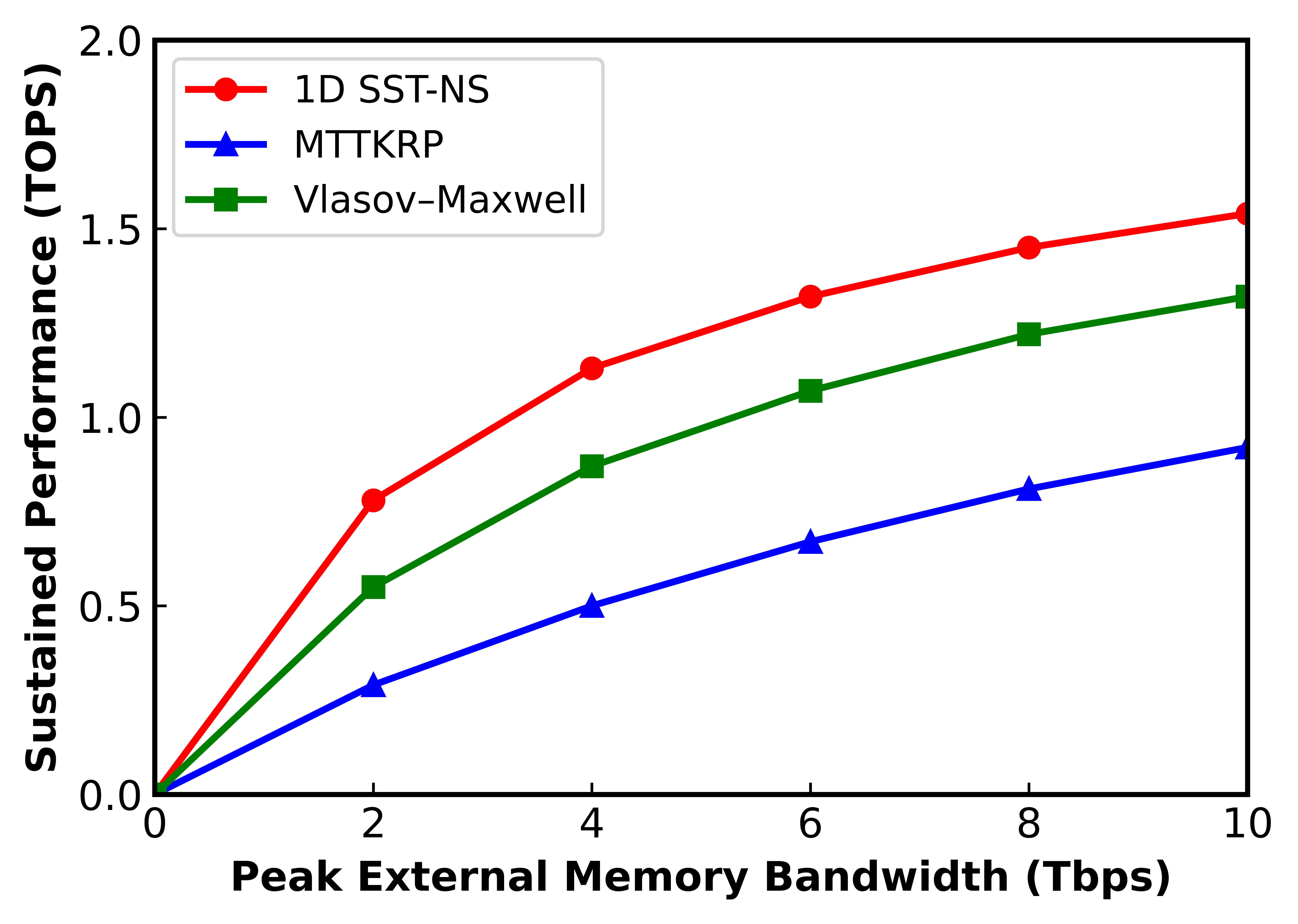}
\vspace{-3mm}
\caption{Impact of peak external memory bandwidth on system performance.}
\label{fig:bw}
\vspace{-1mm}
\end{figure}

Figure~\ref{fig:bw} shows that the sustained performance improves significantly with increased memory bandwidth for all three workloads, highlighting the sensitivity of the streaming algorithm to memory bandwidth constraints. With high-speed memory such as HBM3E (9.8 Tbps)~\cite{10536972}, the workloads achieve substantial performance gains. To further approach peak performance, architectural techniques such as on-chip caching can be employed.

\subsubsection{pSRAM Operating Frequency}
\begin{figure}[!ht]
\vspace{-3mm}
\centering
\includegraphics[width=0.9\linewidth]{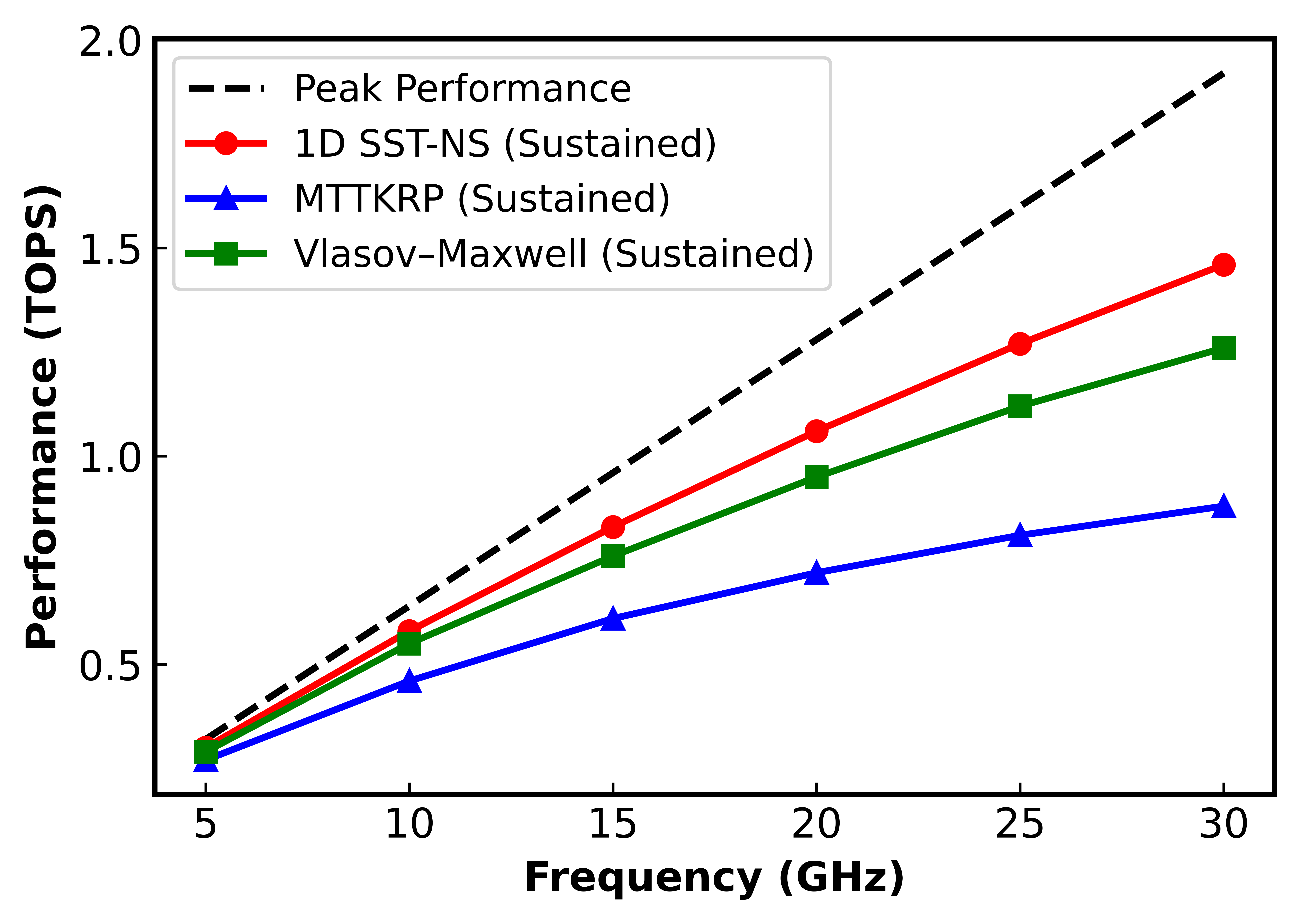}
\caption{Impact of pSRAM operating frequency on system performance.}
\label{fig:freq}
\vspace{-2mm}
\end{figure}

The effect of pSRAM operating frequency on system performance is illustrated in Figure~\ref{fig:freq}. The sustained performance of compute-bound workloads increases nearly linearly with frequency, indicating that higher clock rates directly improve the compute power of the pSRAM array. The increasing gap between peak and sustained performance at higher frequencies highlights the need for architectural co-design of memory and I/O subsystems with high-frequency pSRAM array.

\vspace{2mm}
\subsubsection{Opto-Electronic Conversion Latency}
\begin{figure}[!ht]
\vspace{-3mm}
\centering
\includegraphics[width=0.9\linewidth]{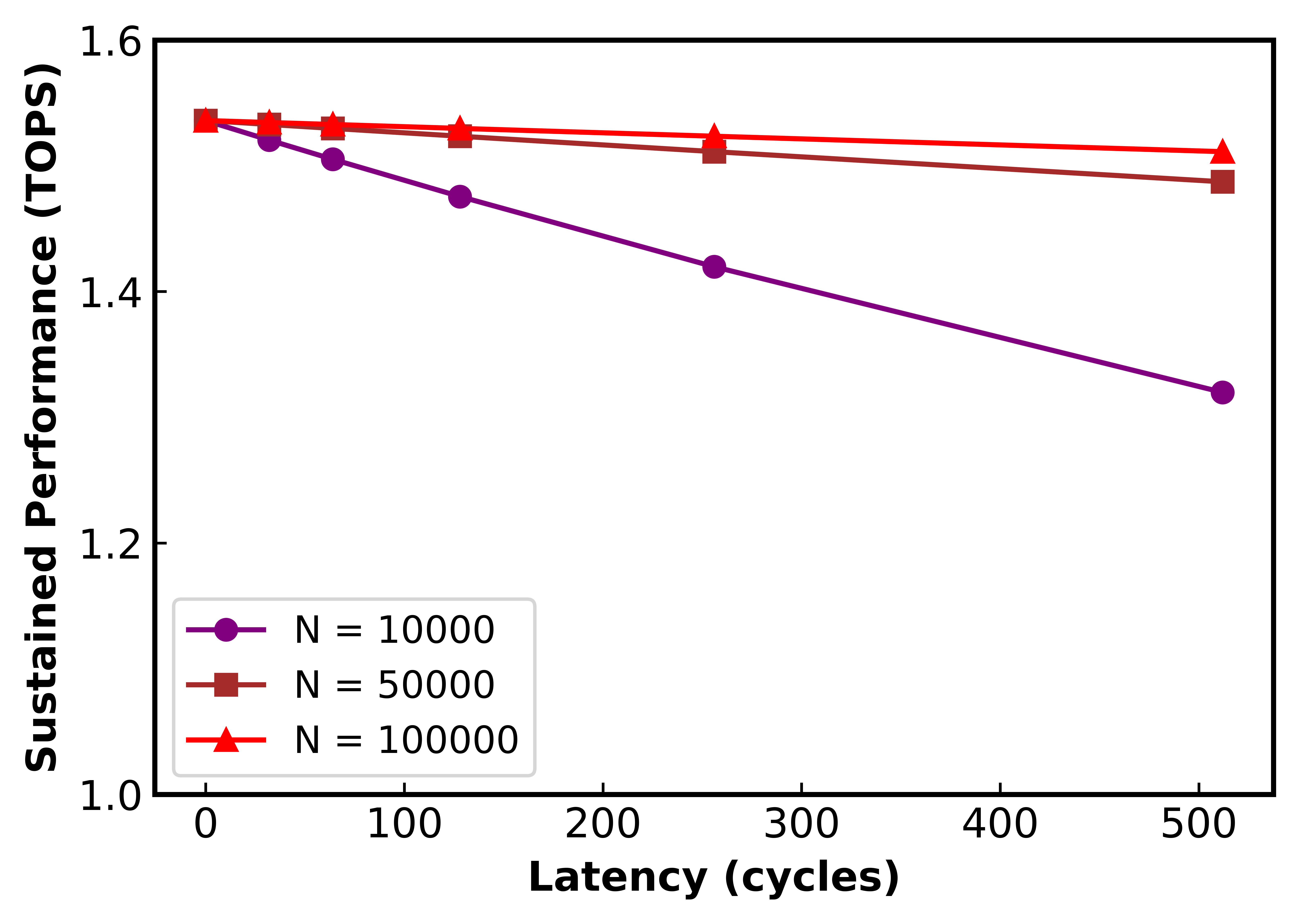}
\caption{Impact of opto-electronic conversion latency on system performance for 1D SST-NS.}
\label{fig:L}
\vspace{-2mm}
\end{figure}

This analysis, as shown in Figure~\ref{fig:L}, focuses on the 1D SST-NS workload with varying grid points $N$, to explore how latency scales with problem size. Our performance model incorporates pipelined opto-electronic conversion in both directions, reflecting a realistic system design. When $N$ is large (e.g., $10000 - 100000$), the impact of conversion latency becomes less significant, as it is amortized over a large number of computations.

\subsection{Scalability with Energy and Area}
At 20 GHz, the pSRAM bitcell exhibits an energy dissipation of 0.5 pJ/bit with two operations (multiply and accumulate) performed per bit, based on device-level measurements~\cite{psram_dac}. Under constant voltage operation, we extrapolate energy to scale linearly with frequency. Table~\ref{tab:energy_efficiency} summarizes the estimated energy dissipation per bit and corresponding energy efficiency (TOPS/W) across multiple operating frequencies, guiding performance-driven design. Each pSRAM bitcell occupies approximately 0.1 $\text{mm}^2$~\cite{psram}, resulting in a total area of 25.6 $\text{mm}^2$ for a 1x256 bit array.
\begin{table}[ht]
\centering
\caption{Estimated Energy Dissipation and Efficiency}
\label{tab:energy_efficiency}
\begin{tabular}{ccc}
\hline
\textbf{Frequency (GHz)} & \textbf{Energy/bit (pJ)} & \textbf{Energy Efficiency (TOPS/W)} \\
\hline
16 & 0.40 & 5.00 \\
20 & 0.50 & 4.00 \\
32 & 0.80 & 2.50 \\
48 & 1.20 & 1.67 \\
\hline
\end{tabular}
\vspace{-2mm}
\end{table}
\vspace{2mm}
\subsection{Scalability with pSRAM Array Size}
\begin{figure}[ht]
\vspace{-2mm}
\centering
\includegraphics[width=0.9\linewidth]{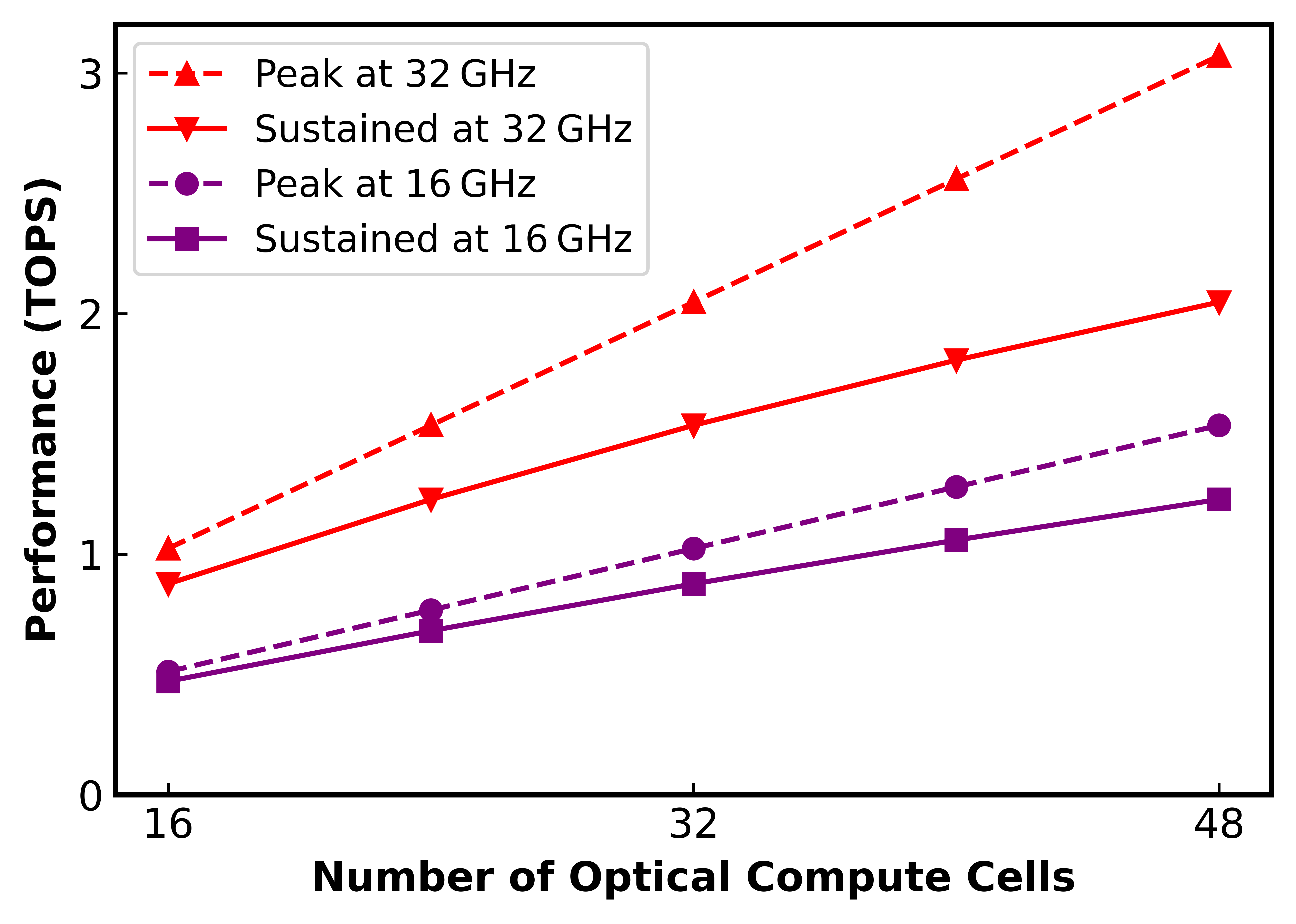}
\caption{Projected performance scaling of pSRAM array with number of optical compute cells for 1D SST-NS.}
\label{fig:P}
\vspace{-3mm}
\end{figure}

Figure~\ref{fig:P} illustrates the projected scaling behavior of the pSRAM array with increasing optical compute cells. Both peak and sustained performance scale with array size at two operating frequencies, confirming architectural scalability. The trend also highlights 
bandwidth-limited saturation at 32 GHz with larger arrays. While we evaluate a pSRAM with 32 compute cells, the result suggests that significantly higher throughput can be achieved with larger arrays by scaling memory bandwidth appropriately.

%% file: 7_Conclusion.tex
\section{Conclusion}~\label{conclusion}
This work presents a comprehensive system-level performance model for pSRAM-based in-memory computing, capturing and analyzing key latency components, including external memory bandwidth and opto-electronic conversion. To enable systematic algorithm-to-hardware mapping, we defined a network model abstraction and designed streaming algorithms for three representative workloads: the Sod shock tube problem, MTTKRP, and the Vlasov–Maxwell equation. We demonstrated that, while accounting for system overheads, a compact 1$\times$256 bit single-wavelength pSRAM array achieves up to 1.5 TOPS sustained throughput at an energy efficiency of 2.5 TOPS/W. 